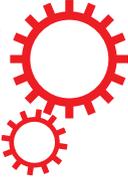

# OPEN  Two-copy Quantum Teleportation

Quan Quan[1,2], Ming-Jing Zhao[3], Shao-Ming Fei[4,5], Heng Fan[6,7,2], Wen-Li Yang[2,8] & Gui-Lu Long[1]

We investigate two-copy scenario of quantum teleportation based on Bell measurements. The detailed protocol is presented and the general expression of the corresponding optimal teleportation fidelity is derived, which is given by the two-copy fully entangled fraction that is invariant under local unitary transformations. We prove that under a specific case of the protocol, which is significant for improving the optimal fidelity, the set of states with their two-copy fully entangled fractions bounded by a threshold value that required for useful two-copy teleportation is convex and compact. Hence the witness operators exist to separate states that are useful for two-copy teleportation from the rest ones. Moreover, we show that the optimal fidelity of two-copy teleportation surpasses that of the original one copy teleportation.

Quantum teleportation plays an important role in quantum information processing[1,2]. It gives ways to transmit an unknown quantum state from a sender traditionally named "Alice" to a receiver "Bob" who are spatially separated, using classical communication and quantum resources[3–6]. In[7,8], the authors consider the original one copy teleportation: Alice and Bob previously share a pair of particles in an arbitrary mixed entangled state $\chi$. In order to teleport an unknown state to Bob, Alice first performs joint Bell measurement on her particles and tells the results to Bob by classical communication. Bob tries his best to choose particular unitary transformations to get the optimal transmission fidelity. The optimal transmission fidelity of such teleportation is given by the fully entangled fraction (FEF)[9] of the quantum resource. It shows that when the resource $\chi$ is a maximally entangled pure state, the corresponding optimal fidelity is equal to 1. However, Alice and Bob usually share a mixed entangled state due to decoherence, and the optimal fidelity is less than 1.

One way to improve the fidelity of teleportation is to distill entanglement[10], which refers to the procedure of converting mixed entangled states into singlets by using many copies of the entangled resources. The distillation of pure states is often referred to as entanglement concentration[11]. For mixed states, since the distillation protocol presented in[10], fruitful results have been obtained[12–15]. However, the problem of distillation is that the complicated protocol may have to be repeated for infinitely many times to bring out a singlet. Moreover, in each round the desired results are usually obtained probabilistically, usually with an extremely low possibility to get an expected measurement outcome.

Inspired by this, to improve the teleportation fidelity, we propose two-copy quantum teleportation scenario directly, instead of bringing out a singlet as the resource of traditional one copy teleportation scenario. Specifically, we introduce a quantum teleportation protocol based on Bell measurements. The corresponding optimal teleportation fidelity is derived analytically. The fact that the optimal fidelity of the two-copy teleportation can surpass that in one copy scenario (the traditional quantum teleportation) is shown by analytical means together with numerical methods. In particular, we discuss a specific case which is significant for improving the optimal fidelity. It shows that the set of quantum states with their two-copy fully entangled fraction bounded by a threshold value required for useful two-copy teleportation is convex and compact, which demonstrates the existence of teleportation witness[16–18] in two-copy quantum teleportation.

## Results

The two-copy teleportation protocol based on Bell measurements is as follows. Initially Alice and Bob share two pairs of entangled resources, see Fig. 1. Particles 1 and 2 (resp. 3 and 4) are in an entangled state $\chi$. Particles 1 and 3 are in Alice's side, while particles 2 and 4 are in Bob's side. Alice wants to transmit an unknown state $\rho_{in}$ of particle 0 to Bob. Firstly, Alice (resp. Bob) performs a joint local unitary operation $W$ (resp. $V$) on particles 1 (resp. 2) and particle 3 (resp. 4). Then she makes joint Bell measurement on particles 0 and 1. She informs Bob the measurement results by classical means. According to these measurement results, Bob chooses corresponding unitary transformations $\{T\}$ on particle 2 to achieve the optimal teleportation fidelity.

[1]State Key Laboratory of Low-Dimensional Quantum Physics and Department of Physics, Tsinghua University, Beijing, 100084, China. [2]Institute of Modern Physics, Northwest University, Xi'an, 710069, China. [3]School of Science, Beijing Information Science and Technology University, Beijing, 100192, China. [4]School of Mathematical Sciences, Capital Normal University, Beijing, 100048, China. [5]Max-Planck-Institute for Mathematics in the Science, Leipzig, 04103, Germany. [6]Institute of Physics, Chinese Academy of Sciences, Beijing, 100190, China. [7]Collaborative Innovation Center of Quantum Matter, Beijing, 100190, China. [8]Shaanxi Key Laboratory for Theoretical Physics Frontiers, Xi'an, 710069, China. Correspondence and requests for materials should be addressed to Q.Q. (email: quanq@mail.tsinghua.edu.cn) or S.-M.F. (email: feishm@mail.cnu.edu.cn) or W.-L.Y. (email: wlyang@nwu.edu.cn)

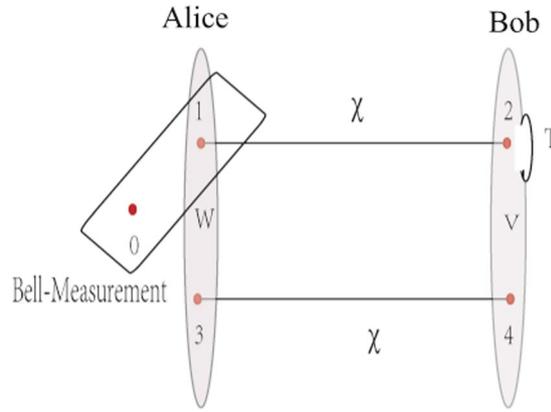

**Figure 1.** Scheme of two-copy teleportation protocol based on Bell measurements. Alice and Bob share two copies of entangled resource $\chi$, with particles 1 and 3 in Alice's side, and particles 2 and 4 in Bob's side. Alice wants to transmit the unknown state $\rho_{in}$ of particle 0 to Bob with optimal fidelity. The two-copy teleportation protocol based on Bell measurements is as follows: firstly, Alice (resp. Bob) performs a joint local unitary operation $W$ (resp. $V$) on particles 1 (resp. 2) and particle 3 (resp. 4) to correlate these two particles. Then Alice makes joint Bell measurement on particles 0 and 1 and informs Bob the measurement results by classical communication. According to the measurement results, Bob chooses corresponding unitary transformations $\{T\}$ on his particles 2 and 4 to restore the input state $\rho_{in}$ on particle 2.

Let $H$ denote an $n$-dimensional Hilbert space, with $\{|j\rangle, j = 0, ..., n-1, n < \infty\}$ as orthogonal normalized basis. A set of unitary matrices $\{U_{st}\}$ in $H$ can be written as: $U_{st} = h^t g^s$, where $h$ and $g$ are $n \times n$ matrices such that $h|k\rangle = |(k+1)/\mod n\rangle$ and $g|k\rangle = w^k|k\rangle$, with $w = \exp\{-2i\pi/n\}$. $\{U_{st}\}$ has the following relations[19]: $\mathrm{tr}(U_{st}^\dagger U_{s't'}) = n\delta_{tt'}\delta_{ss'}$, $U_{st}U_{st}^\dagger = I_{n \times n}$, where $I$ is the identity matrix. The generalized Bell states[7] are given by $|\Phi_{st}\rangle = (I_{n \times n} \otimes U_{st})|\Phi\rangle$, where $|\Phi\rangle = |\Phi_{00}\rangle = \frac{1}{\sqrt{n}}\sum_{j=0}^{n-1}|jj\rangle$ is the maximally entangled pure state. The $n^2$ generalized Bell states $\{|\Phi_{st}\rangle\} = \frac{1}{\sqrt{n}}\sum_{j,k}(U_{st})^*_{jk}|jk\rangle$ form a complete orthogonal normalized basis of the $H \otimes H$ space. Throughout this paper we adopt the standard notations: for any matrix $A \in \mathrm{End}(H)$, $A_\alpha$ is an embedded operator in the tensor space $H \otimes H \otimes \ldots \otimes H$, which acts as $A$ on the $\alpha$-th space and as identity on the other spaces; and for any matrix $U \in \mathrm{End}(H \otimes H)$, $U_{\alpha\beta}$ is an embedded operator in $H \otimes H \otimes \ldots \otimes H$, which acts as identity on the spaces except for the $\alpha$-th and $\beta$-th ones. After some tedious calculation, we get the output state under the scenario of two-copy teleportation:

**Lemma 1** For any unknown input state $\rho_{in}$, the two-copy teleportation protocol under Bell measurements maps the state $\rho_{in}$ to state $\Lambda_{\chi^{\otimes 2}}^{\{T_{st}, W, V\}}(\rho_{in})$,

$$\Lambda_{\chi^{\otimes 2}}^{\{T_{st}, W, V\}}(\rho_{in}) = \frac{1}{n^3} \sum_{s_1,t_1 s'_1 t'_1 s_2, t_2 s'_2 t'_2} \sum_{s,t} \langle \Phi_{s_1 t_1}|\chi|\Phi_{s'_1 t'_1}\rangle \langle \Phi_{s_2 t_2}|\chi|\Phi_{s'_2 t'_2}\rangle$$
$$\times \mathrm{tr}_4[(T_{st})_2 V_{24}(U_{s_1 t_1})_2 (U_{s_2 t_2})_4 W_{24}(U_{st})_2^\dagger(\rho_{in})_2$$
$$\times (U_{st})_2 W_{24}^\dagger(U_{s'_1 t'_1}^\dagger)_2 (U_{s'_2 t'_2}^\dagger)_4 V_{24}^\dagger(T_{st})_2^\dagger], \quad (1)$$

where $W$ and $V$ are the unitary transformations Alice and Bob apply to their two particles, respectively. $T_{st} \in \{T\}$ is the unitary operator that Bob performs on particle 2 to achieve the optimal teleportation fidelity.

**Proof.** First consider that the unknown initial input state $\rho_{in}$ that Alice wants to teleport is a pure state, $|\phi\rangle = \sum_\nu \alpha_\nu|\nu\rangle$.

1). The two entangled resource states shared by Alice and Bob are pure: $\chi^{\otimes 2} = |\Psi\rangle_{1234}\langle\Psi|$, where $|\Psi\rangle_{1234} = \sum_{j,k=0}^{n-1}\sum_{l,m=0}^{n-1} a_{jk}|jk\rangle \otimes a_{lm}|lm\rangle$, $\sum_{j,k=0}^{n-1}|a_{jk}|^2 = 1$. Alice and Bob apply the unitary transformations $W$ and $V$ to their two resource particles respectively. Before the measurement, the initial state becomes $|\phi\rangle_0 W_{13}V_{24}|\Psi\rangle_{1234} = \sum_{j,k=0}^{n-1}\sum_{l,m=0}^{n-1}\sum_{j',k'}^{n-1}\sum_{l',m'}^{n-1}\sum_\nu^{n-1} a_{jk}a_{lm}W_{j'l'}^{jl}V_{k'm'}^{km}\alpha_\nu|\nu j'k'l'm'\rangle_{01234}$. After Alice's joint Bell measurement based on $|\Phi_{st}\rangle$ on particles 0 and 1, we get: $\langle\Phi_{st}|_{01}(|\phi\rangle_0 W_{13}V_{24}|\Psi\rangle_{1234}) = V_{24}A_2 A_4 W_{24}(U_{st}^\dagger)_2|\phi\rangle_2|\Phi\rangle_{34}$, where $A$ is the $n \times n$ matrix with elements $(A)_{jk} = a_{jk}$. Receiving Alice's measurement outcomes, correspondingly Bob applies unitary operators $\{T\}$ on particle 2. The resulting state becomes $(T_{st})_2 V_{24}A_2 A_4 W_{24}(U_{st}^\dagger)_2|\phi\rangle_2|\Phi\rangle_{34}$. Taking partial trace over the spaces with respect to particles 3 and 4, we have

$$\Lambda_{\chi^{\otimes 2}}^{\{T_{st}, W, V\}}(\rho_{in}) = \sum_{s,t}\frac{1}{n}\mathrm{tr}_4[(T_{st})_2 V_{24}A_2 A_4 W_{24}(U_{st}^\dagger)_2|\phi\rangle_2\langle\phi|_2(U_{st})_2 W_{24}^\dagger A_2^\dagger A_4^\dagger V_{24}^\dagger(T_{st})_2^\dagger], \quad (2)$$

which is equivalent to Eq. (1).

2). Now consider the case of arbitrary entangled mixed resources, $\chi^{\otimes 2} = \sum_{\alpha,\beta} P_\alpha P_\beta |\Psi_{\alpha\beta}\rangle\langle\Psi_{\alpha\beta}|$, where $|\Psi_{\alpha\beta}\rangle = \sum_{j,k=0}^{n-1}\sum_{l,m=0}^{n-1} a_{jk}^{(\alpha)}|jk\rangle \otimes a_{lm}^{(\beta)}|lm\rangle$, $0 \leq P_{\alpha(\beta)} \leq 1$ and $\sum_{\alpha(\beta)} P_{\alpha(\beta)} = 1$. Similar to the derivation of (2), we have

$$\Lambda_{\chi^{\otimes 2}}^{\{T_{st},W,V\}}(\rho_{in}) = \frac{1}{n}\sum_{s,t}\sum_{\alpha,\beta} P_\alpha P_\beta \mathrm{tr}_4[(T_{st})_2 V_{24} A_2^{(\alpha)} A_4^{(\beta)} W_{24}(U_{st}^\dagger)_2|\phi\rangle_2\langle\phi|_2 (U_{st})_2 W_{24}^\dagger A_2^{(\alpha)\dagger} A_4^{(\beta)\dagger} V_{24}^\dagger (T_{st})_2^\dagger],$$

where $(A)_{jk}^{(\alpha/\beta)} = a_{jk}^{(\alpha/\beta)}$. Since each matrix $A^{(\alpha)}$ can be decomposed in the basis of $U_{st}$: $A^{(\alpha)} = \sum_{s,t} a_{st}^{(\alpha)} U_{st}$, by using the relation[7], $n\sum_\alpha p_\alpha a_{st}^{(\alpha)} a_{s't'}^{(\alpha)*} = \langle\Phi_{st}|\chi|\Phi_{s't'}\rangle$, we can straightforwardly show that Eq. (1) is also valid for any mixed input state $\rho_{in}$. □

**Remark** The two-copy teleportation scenario is trace preserving, $tr[\Lambda_{\chi^{\otimes 2}}^{\{T_{st},W,V\}}(\rho_{in})] = 1$, see proof in Method.

Utilizing the output state, we get the optimal teleportation fidelity (see Method for detailed proof):

**Theorem 1** The optimal teleportation fidelity $f_2(\chi)_{\max}$ of the two-copy teleportation protocol is given by

$$f_2(\chi)_{\max} = \frac{nF_2(\chi)}{(n+1)} + \frac{1}{n+1}, \tag{3}$$

where $\Omega_{13}^T = W_{13}(U_{st})_1 (T_{st})_3$, $F_2(\chi)$ is two-copy fully entangled fraction,

$$F_2(\chi) = \max_{\Omega,V \in U(n^2)} \{\langle\Phi|_{12} \mathrm{tr}_{34}[\Omega_{13} V_{24} \chi_{12} \chi_{34} \Omega_{13}^\dagger V_{24}^\dagger]|\Phi\rangle_{12}\}.$$

From Theorem 1 we see that the two-copy optimal teleportation fidelity $f_2(\chi)$ solely depends on the two-copy fully entangled fraction $F_2(\chi)$. It can be shown that $F_2(\chi)$ given by (3) is an invariant under local unitary transformations: $\chi_{12}\chi_{34} \to (\mathfrak{U})_{13}(\mathfrak{V})_{24}\chi_{12}\chi_{34}(\mathfrak{U})_{13}^\dagger(\mathfrak{V})_{24}^\dagger$, where $\mathfrak{U}$ and $\mathfrak{V}$ are unitary operators on $H \otimes H$. Theorem 1 also tells us that a resource state $\chi$ is useful, namely, it gives better teleportation fidelity than classical channels, if $F_2(\chi) > \frac{1}{n}$.

The original one copy optimal teleportation fidelity is given by[7,8]

$$f_1(\chi)_{\max} = \frac{nF_1(\chi)}{n+1} + \frac{1}{n+1}, \tag{4}$$

where $F_1(\chi) = \max_{U \in U(n)}\{\langle\Phi|_{12} U_2^\dagger \chi_{12} U_2|\Phi\rangle_{12}\}$ is the original fully entangled fraction. To show that the two-copy teleportation protocol is always better, or at least as good as the original one copy case, let us simply choose the unitary matrix $W_{13} = V_{24} = I_{n^2 \times n^2}$ in Theorem 1, which does not necessarily reach the value of $f_2(\chi)_{\max}$.

**Lemma 2** When choose the unitary matrix $W_{13} = V_{24} = I_{n^2 \times n^2}$ in Theorem 1, the output state $\Lambda_{\chi^{\otimes 2}}^{\{T_{st},W,V\}}(\rho_{in})$ of the two-copy teleportation protocol reduces to the output state $\Lambda^{(\chi)}(\{T\})(\rho)$ of one-copy teleportation protocol[7,8] (see Method).

**Theorem 2** The two-copy optimal teleportation fidelity is always greater than or equal to that of the original one copy protocol, that is, for any arbitrary state $\chi$,

$$f_2(\chi)_{max} \geq f_1(\chi)_{max}. \tag{5}$$

*Proof.* From Eqs (3) and (4), one can see that both optimal teleportation fidelities for two-copy and onecopy teleportation protocols are linear functions of the corresponding fully entangled fractions. These fully entangled fractions characterize the usefulness of the entangled resource states in quantum teleportation. To compare $f_2(\chi)_{max}$ with $f_1(\chi)_{max}$, one only needs to compare $F_2(\chi)$ with $F_1(\chi)$. Unfortunately, both $F_2(\chi)$ and $F_1(\chi)$ are formidably difficult to calculate analytically. Analytical formulae for $F_1(\chi)$ are only available for some special states[20,21]. Generally one has only estimations of the upper and lower bounds of $F_1(\chi)$[21,22]. The computation of $F_2(\chi)$ is much more difficult than that of $F_1(\chi)$. However, if one takes $W = V$ to be identity, or takes $\Omega$ and $V$ in (3) to be the tensor of two unitary operators $Y \otimes \Gamma$ with $Y, \Gamma \in U(n)$, then one gets $F_2(\chi) = F_1(\chi)$. Thus the extreme value range of $F_2$ is larger than that of $F_1$. Therefore, for any arbitrary state $\chi$, $f_2(\chi)_{max} \geq f_1(\chi)_{max}$, i.e., the two-copy optimal teleportation fidelity is always greater than or equal to that of the original one copy protocol. □

In the following, we give numerical calculations of $F_2$ and $F_1$ by using the Conjugate Gradient Algorithm[23,24]. Following the modified Polak-Ribiere-Polyak method introduced in[24], we can get the numerical result of $F_1$.

**Lemma 3** To simplify the computation, we take $V = I_{n \times n}$ to get a lower bound $F'_2(\chi)$ of $F_2(\chi)$:

$$F'_2(\chi) = \max_{\Omega \in U(n^2)}\left\{\sum_j \langle\Phi|_{12}\langle j|_3 \Omega_{23}\chi_{12}\Omega_{23}^\dagger \rho_3^*|j\rangle_3|\Phi\rangle_{12}\right\},$$

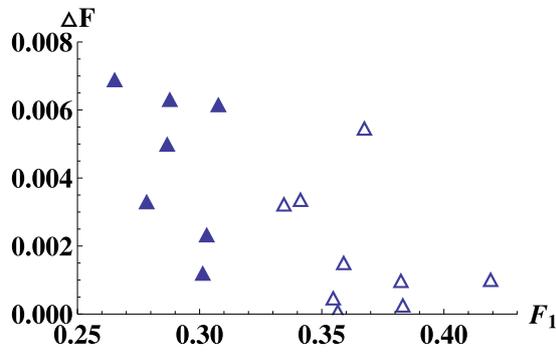

**Figure 2.** Hollow (solid) triangles stand for 3 (4)-dimensional randomly generated states. Horizontal axis is the one-copy fully entangled fraction $F_1$. Vertical axis denotes the difference $\Delta F = F'_2 - F_1$. It is seen that $\Delta F > 0$, and hence the lower bound of the optimal fidelity of the two-copy teleportation is better than the optimal fidelity of original one-copy teleportation for all these randomly generated states.

where $\rho_3 = tr_4(\chi_{34})$; $\rho_3^*$ is the conjugate of $\rho_3$, see Method.

Denote $\mathfrak{F}_2(\chi) = \sum_j \langle\Phi|_{12} \langle j|_3 \Omega_{23}\chi_{12}\Omega_{23}^\dagger \rho_3^*|j\rangle_3 |\Phi\rangle_{12}$. We get $F'_2(\chi) = \max_{\Omega_{23} \in U(n^2)} \mathfrak{F}_2$. Set $\Delta F = F'_2(\chi) - F_1(\chi)$. Figure 2 shows that for these randomly generated states, one has $F'_2(\chi) > F_1(\chi)$, i.e., the lower bound of the optimal fidelity of two-copy teleportation is better than the optimal fidelity of the one-copy teleportation.

Let us investigate further the lower bound $F'_2(\chi)$ of the two-copy fully entangled fraction $F_2(\chi)$, obtained by setting $V = I_{n \times n}$ in the two-copy teleportation protocol. For the original one copy teleportation, it is shown that the resource states $\chi$ satisfying $F_1(\chi) > 1/n$ are useful for teleportation[17]. Since the set of states satisfying $F_1(\chi) \leq \frac{1}{n}$ is convex and compact, there exist witness operators that detect entangled states which are useful for teleportation[17,18]. Similarly, for the two-copy teleportation protocol with $V = I_{n \times n}$, if resource states satisfying $F'_2(\chi) > \frac{1}{n}$, they are useful for teleportation. Denote $\mathbb{S} = \left\{\chi: F'_2(\chi) \leq \frac{1}{n}\right\}$. We have:

**Theorem 3** The set $\mathbb{S}$ is convex and compact.

**Proof.**

1). The set $\mathbb{S}$ is convex: let $\chi_a$ and $\chi_b \in \mathbb{S}$, namely, $F'_2(\chi_a) \leq \frac{1}{n}, F'_2(\chi_b) \leq \frac{1}{n}$. Consider $\chi_c = \xi\chi_a + (1-\xi)\chi_b$, where $\xi \in [0, 1]$. By the definition of $F'_2(\chi) = \max_{\Omega_{23} \in U(n^2)}\left\{\sum_j \langle\Phi_{00}|_{12} \langle j|_3 \Omega_{23}\chi_{12}\Omega_{23}^\dagger \rho_3^*|j\rangle_3 |\Phi_{00}\rangle_{12}\right\}$, we get that $F'_2(\chi_c) \leq \xi F'_2(\chi_a) + (1 - \xi)F'_2(\chi_b) \leq \frac{1}{n}$. Thus $\chi_c \in \mathbb{S}$, i.e. $\mathbb{S}$ is convex.

2). The set $\mathbb{S}$ is compact: for finite dimensional Hilbert spaces, to show that a set is compact, it is enough to show the set is closed and bounded. $\mathbb{S}$ is bounded since the eigenvalues of $\chi$ lies in $[0, 1]$[17]. To see that it is closed, assume that for any two density matrices $\chi_a$ and $\chi_b$, the value of $F'_2(\chi_a + \chi_b)$ and $F'_2(\chi_a)$ are obtained at $\Omega_{a+b}$ and $\Omega_a$ respectively, where $\Omega \in U(n^2)$. Therefore

$$F'_2(\chi_a + \chi_b) - F'_2(\chi_a) \leq n^2 ||\Omega_{a+b}||^2 (2||\chi_a|| + ||\chi_b||)||\chi_b||, \quad (6)$$

where $||\chi_a||$ is the maximal eigenvalue of $\chi_a$ satisfying $||\chi_a|| \leq 1$. Since the set of all unitary operators is bounded, $||\Omega_{a+b}||^2 \leq \nu$, where $\nu$ is a positive real number. Thus $F'_2(\chi_a + \chi_b) - F'_2(\chi_a) \leq n^2\nu(2 + ||\chi_b||)||\chi_b||$. □

**Remark** For the two-copy teleportation protocol with $V = I_{n \times n}$, there exist witnesses to identify the usefulness of an unknown resource state experimentally.

In fact, according to the Hahn-Banach theorem[25], any $\chi \notin \mathbb{S}$ can be separated from $\mathbb{S}$ by a hyperplane. This feature enables for the existence of hermitian witness operators and thus experimental ways to detect the usefulness of an unknown resource state. □

## Discussion

To conclude, we have proposed a general two-copy quantum teleportation protocol based on Bell measurement systematically. The corresponding optimal teleportation fidelity have been analytically derived. Interestingly, the formulae of optimal teleportation fidelity in two-copy scenario and one-copy scenario are similar. Both of them are the one-variable linear function of the corresponding fully entangled fractions, which are invariants under local unitary transformations on the resource states. Analytical analysis together with numerical results illustrate that the optimal teleportation fidelity can be improved in two-copy teleportation protocol when compared with the onecopy scenario. Therefore, in order to improve the teleportation fidelity, two-copy teleportation protocol is significant both theoretically and experimentally. Furthermore, we have shown that in the context of two-copy teleportation protocol, if one considers a specific case that Bob conducts identity transformation on his resource states, the optimal fidelity can still be improved. The set of quantum states with their two-copy fully entangled fraction bounded by a threshold value required for useful two-copy teleportation is convex and compact. It demonstrates the existence of two-copy teleportation witness, completing the theory of two-copy quantum teleportation. Besides, this two-copy quantum teleportation protocol can be generalized to many-copy cases and may result in further improvement on the teleportation fidelity. Different protocols and methods are worthwhile to conceive and investigate in the future.

## Methods

**Proof of Remark.** The output state $\Lambda_{\chi^{\otimes 2}}^{\{T_{st}\},W,V\}}$ of the two-copy teleportation is trace preserving:

$$\begin{aligned}\operatorname{tr}[\Lambda_{\chi^{\otimes 2}}^{\{T_{st}\},W,V\}}(\rho_{in})] &= \frac{1}{n^3}\sum_{s_1,t_1 s'_1 t'_1 s_2, s'_2 t'_2}\sum\sum\sum \langle\Phi_{s_1 t_1}|\chi|\Phi_{s'_1 t'_1}\rangle \\ &\quad \times \langle\Phi_{s_2 t_2}|\chi|\Phi_{s'_2 t'_2}\rangle \operatorname{tr}_2\{(T_{st})_2(U_{s_1 t_1})_2 \operatorname{tr}_4[(U_{s_2 t_2})_4 W_{24} W_{24}^\dagger (U_{s'_2 t'_2})_4^\dagger](U_{s'_1 t'_1}^\dagger)_2 (T_{st})_2^\dagger\}\\ &= \frac{1}{n^2}\sum_{s_1,t_1 s'_1 t'_1 s_2, s'_2 t'_2}\sum\sum\sum \langle\Phi_{s_1 t_1}|\chi|\Phi_{s'_1 t'_1}\rangle\\ &\quad \langle\Phi_{s_2 t_2}|\chi|\Phi_{s'_2 t'_2}\rangle \operatorname{tr}_2[(U_{s_1 t_1})_2 (U_{s'_1 t'_1}^\dagger)_2]\operatorname{tr}_4[(U_{s_2 t_2})_4 (U_{s'_2 t'_2}^\dagger)_4] = 1,\end{aligned}$$

where in the first equality we have used the relation $\sum_{s,t} U_{st}^\dagger A U_{st} = n\operatorname{tr}(A)I_{n\times n}$ for any $n\times n$ matrix $A$. □

**Proof of Theorem 1.** Let $U(n)$ be an irreducible n-dimensional representation of unitary group G. By using the Schur's lemma

$$\int_G dg (U^\dagger(g)\otimes U^\dagger(g))\sigma(U(g)\otimes U(g)) = \alpha_1 I\otimes I + \alpha_2 P,$$

$$\alpha_1 = \frac{n^2 \operatorname{tr}(\sigma) - n\operatorname{tr}(\sigma P)}{n^2(n^2-1)}, \quad \alpha_2 = \frac{n^2 \operatorname{tr}(\sigma P) - n\operatorname{tr}(\sigma)}{n^2(n^2-1)},$$

where $\sigma$ is any operator acting on the tensor space, $P$ is the flip operator, $dg$ is the Haar measure on $G$ normalized by $\int_G dg = 1$, we get the fidelity of the two-copy teleportation protocol,

$$\begin{aligned}f_2(\chi) &= \overline{\langle\phi_{in}|\Lambda_{\chi^{\otimes 2}}^{\{T_{st}\},W,V\}}(\rho_{in})|\phi_{in}\rangle}\\ &= \frac{1}{n^3}\sum_{s_1,t_1 s'_1,t'_1 s_2,s'_2,t'_2}\sum\sum\sum \langle\Phi_{s_1 t_1}|\chi|\Phi_{s'_1 t'_1}\rangle\langle\Phi_{s_2 t_2}|\chi|\Phi_{s'_2 t'_2}\rangle\\ &\quad \times \sum_{s,t,j,k}\langle 00|\int_G [U(g)^\dagger \otimes U(g)^\dagger][\langle j|_4\\ &\quad \times (T_{st})_2 V_{24}(U_{s_1 t_1})_2(U_{s_2 t_2})_4 W_{24}(U_{st}^\dagger)_2|k\rangle_4]\\ &\quad \otimes [\langle k|_4 (U_{st})_2 W_{24}^\dagger (U_{s'_1 t'_1}^\dagger)_2(U_{s'_2 t'_2}^\dagger)_4 V_{24}^\dagger\\ &\quad \times (T_{st})_2^\dagger |j\rangle_4][U(g)\otimes U(g)]dg|00\rangle\\ &= \frac{1}{n^4(n+1)}\sum_{s_1,t_1 s'_1,t'_1 s_2,s'_2,t'_2}\sum\sum\sum \langle\Phi_{s_1 t_1}|\Phi_{s'_1 t'_1}\rangle\langle\Phi_{s_2 t_2}|\Phi_{s'_2 t'_2}\rangle\\ &\quad \times \sum_{s,t,j,k,l,l'}\{\operatorname{tr}_2[\langle j|_4 (T_{st})_2 V_{24}(U_{s_1 t_1})_2(U_{s_2 t_2})_4|l\rangle_4 \langle l|_4\\ &\quad \times W_{24}(U_{st}^\dagger)_2|k\rangle_4]\operatorname{tr}_4[\langle k|_4 (U_{st})_2 W_{24}^\dagger\\ &\quad \times (U_{s'_1 t'_1}^\dagger)_2(U_{s'_2 t'_2}^\dagger)_4|l'\rangle_4 \langle l'|_4 V_{24}^\dagger(T_{st})_2^\dagger|j\rangle_4]\\ &\quad + \frac{1}{n^2(n+1)}\sum_{s_1,t_1 s'_1,t'_1 s_2,s'_2,t'_2}\sum\sum\sum \langle\Phi_{s_1 t_1}|\chi|\Phi_{s'_1 t'_1}\rangle\langle\Phi_{s_2 t_2}|\chi|\Phi_{s'_2 t'_2}\rangle\\ &\quad \times \operatorname{tr}_2[(U_{s_1 t_1})_2(U_{s'_1 t'_1}^\dagger)_2]\operatorname{tr}_4[(U_{s_2 t_2})_4(U_{s'_2 t'_2}^\dagger)_4]\\ &= \frac{1}{n^4(n+1)}\sum_{s_1,t_1 s'_1,t'_1 s_2,s'_2,t'_2,s,t,j,k}\sum \langle\Phi|_{12}\langle\Phi|_{34}\operatorname{tr}_{24}[W_{24}(U_{st}^\dagger)_2|k\rangle_4\\ &\quad \times \langle j|_4 (T_{st})_2 V_{24}(U_{s_1 t_1})_2(U_{s_2 t_2})_4](U_{s_1 t_1}^\dagger)_2(U_{s_2 t_2}^\dagger)_4 \chi_{12}\chi_{34}\\ &\quad \times \operatorname{tr}_{24}[V_{24}^\dagger (T_{st})_2^\dagger|j\rangle_4\langle k|_4 (U_{st})_2 W_{24}^\dagger (U_{s'_1 t'_1}^\dagger)_2(U_{s'_2 t'_2}^\dagger)_4]\\ &\quad \times (U_{s'_1 t'_1})_2(U_{s'_2 t'_2})_4|\Phi\rangle_{12}|\Phi\rangle_{34} + \frac{1}{n+1}\\ &= \frac{1}{(n+1)}\sum_{s,t}\langle\Phi|_{12}\langle\Phi|_{34} W_{24}(U_{st}^\dagger)_2 (T_{st})_2\\ &\quad \times \operatorname{tr}_4[V_{24}\chi_{12}\chi_{34} V_{24}^\dagger](T_{st})_2^\dagger (U_{st})_2 W_{24}^\dagger)|\Phi\rangle_{12}|\Phi\rangle_{34} + \frac{1}{n+1},\end{aligned}$$

where $\overline{\langle\phi_{in}|\ldots|\phi_{in}\rangle}$ represents the average over all input states $|\phi_{in}\rangle$.

Then the optimal teleportation fidelity is given by the maximal fidelity of $f_2(\chi)$,

$$\begin{aligned}f_2(\chi)_{max} &= \frac{n^2}{(n+1)}\max_{\Omega,V\in U(n^2)}\{\langle\Phi|_{12}\langle\Phi|_{34}\Omega_{24}\text{tr}_4[V_{24}\chi_{12}\chi_{34}V_{24}^\dagger]\Omega_{24}^\dagger|\Phi\rangle_{12}|\Phi\rangle_{34}\} + \frac{1}{n+1}\\ &= \frac{n^2}{(n+1)}\max_{\Omega,V\in U(n^2)}\{\langle\Phi|_{12}\langle\Phi|_{34}\Omega_{13}^T\text{tr}_4[V_{24}\chi_{12}\chi_{34}V_{24}^\dagger]\Omega_{13}^*|\Phi\rangle_{12}|\Phi\rangle_{34}\} + \frac{1}{n+1},\end{aligned}$$

where $\Omega_{24} = W_{24}(U_{st})_2(T_{st})_2$. Rewriting $\Omega^T$ as $\Omega$, we get (3). □

**Proof of Lemma 2.** When we choose $W_{13} = V_{24} = I_{n^2\times n^2}$, the output state $\Lambda^{\{T_{st},W,V\}}_{\chi^{\otimes 2}}(\rho_{in})$ of the two-copy teleportation protocol reduces to that of one-copy teleportation protocol $\Lambda^{(\chi)}(\{T\})(\rho)$ in[7]:

$$\begin{aligned}\Lambda^{\{T_{st},W,V\}}_{\chi^{\otimes 2}}(\rho_{in}) &\\ &= \frac{1}{n^3}\sum_{s_1,t_1s'_1t'_1s_2,t_2s'_2t'_2}\sum_{s,t}\langle\Phi_{s_1t_1}|\chi|\Phi_{s'_1t'_1}\rangle\langle\Phi_{s_2t_2}|\chi|\Phi_{s'_2t'_2}\rangle\text{tr}_4[(T_{st})_2(U_{s_1t_1})_2\\ &\quad\times (U_{s_2t_2})_4(U_{st})_2^\dagger|\phi\rangle_2\langle\phi|_2(U_{st})_2(U_{s'_1t'_1}^\dagger)_2(U_{s'_2t'_2})_4^\dagger(T_{st})_2^\dagger]\\ &= \Lambda^{(\chi)}(\{T\})(\rho_{in})\frac{1}{n}\sum_{s_2,t_2s'_2,t'_2}\langle\Phi_{s_2t_2}|\chi|\Phi_{s'_2t'_2}\rangle\text{tr}[U_{s_2t_2}U_{s'_2t'_2}^\dagger] = \Lambda^{(\chi)}(\{T\})(\rho_{in}).\end{aligned}$$

**Proof of Lemma 3**

$$\begin{aligned}F'_2(\chi) &= n\max_{\Omega\in U(n^2)}\{\langle\Phi|_{12}\langle\Phi|_{34}\Omega_{24}\chi_{12}\rho_3\Omega_{24}^\dagger|\Phi\rangle_{12}|\Phi\rangle_{34}\}\\ &= \max_{\Omega\in U(n^2)}\{\sum_{j,j'}\langle\Phi|_{12}\langle j|_4\Omega_{24}\chi_{12}\Omega_{24}^\dagger|j'\rangle_4|\Phi\rangle_{12}\langle j|_3\rho_3|j'\rangle_3\}\\ &= \max_{\Omega\in U(n^2)}\{\sum_j\langle\Phi|_{12}\langle j|_3\Omega_{23}\chi_{12}\Omega_{23}^\dagger\rho_3^*|j\rangle_3|\Phi\rangle_{12}\},\end{aligned}$$

where $\rho_3 = \text{tr}_4(\chi_{34})$.

**Numerical calculation for Fig. 2.** Denote $\mathfrak{F}_1(\chi) = \langle\Phi|_{12}U_2^\dagger\chi_{12}U_2|\Phi\rangle_{12}$. Since unitary $U$ can be expressed as $U = exp\{i*H\}$, where H is the corresponding Hermitian matrix, we can get the increment of $\mathfrak{F}_1$

$$\begin{aligned}\Delta\mathfrak{F}_1 &= \frac{i}{2}[\langle\Phi|_{12}(\Delta\mathbb{H}_2 U_2 + U_2\Delta\mathbb{H}_2)\chi_{12}U_2^\dagger|\Phi\rangle_{12}\\ &\quad - \langle\Phi|_{12}U_2\chi_{12}(\Delta\mathbb{H}_2 U_2^\dagger + U_2^\dagger\Delta\mathbb{H}_2)|\Phi\rangle_{12}]\\ &= \frac{i}{2}\text{tr}_2\{\Delta\mathbb{H}_2[\text{tr}_1(U_2\chi_{12}U_2^\dagger|\Phi\rangle_{12}\langle\Phi|_{12})\\ &\quad + \text{tr}_1(\chi_{12}U_2^\dagger|\Phi\rangle_{12}\langle\Phi|_{12}U_2)\\ &\quad - \text{tr}_1(U_2^\dagger|\Phi\rangle_{12}\langle\Phi|_{12}U_2\chi_{12})\\ &\quad - \text{tr}_1(|\Phi\rangle_{12}\langle\Phi|_{12}U_2\chi_{12}U_2^\dagger)]\}\\ &= \frac{i}{2}\text{tr}_2[\Delta\mathbb{H}_2 G_2(\mathfrak{F}_1)].\end{aligned}$$

Choosing $G_2(\mathfrak{F}_1)$ as the gradient of $\mathfrak{F}_1$, then following the MRPR Method introduced from Eq. (2.1) to Eq. (2.4) in[24], we can get the numerical result of $F_1$.

Using the same method, we can obtain the increment of $\mathfrak{F}'_2(\chi)$:

$$\begin{aligned}\Delta\mathfrak{F}'_2 &= \frac{i}{2}\text{tr}_{23}\{\Delta\mathbb{H}_{23}[\text{tr}_1(\Omega_{23}\chi_{12}\Omega_{23}^\dagger\rho_3^*|\Phi\rangle_{12}\langle\Phi|_{12}) + \text{tr}_1(\chi_{12}\Omega_{23}^\dagger\rho_3^*|\Phi\rangle_{12}\langle\Phi|_{12}\Omega_{23})\\ &\quad - \text{tr}_1(\Omega_{23}^\dagger\rho_3^*|\Phi\rangle_{12}\langle\Phi|_{12}\Omega_{23}\chi_{12}) - \text{tr}_1(\rho_3^*|\Phi\rangle_{12}\langle\Phi|_{12}\Omega_{23}\chi_{12}\Omega_{23}^\dagger)]\}\\ &= \frac{i}{2}\text{tr}_{23}[\Delta\mathbb{H}_{23}G_{23}(\mathfrak{F}'_2)].\end{aligned}$$

Thus we can get a numerical lower bound of $F'_2$. We generate random 3d and 4d states by Mathematica and let $\Delta F = F'_2 - F_1$, then can get into the result of Fig. 2.

**Detailed proof of Theorem 3**

$$F'_2(\chi_a + \chi_b) - F'_2(\chi_a)$$
$$\leq \sum_j \langle \Phi_{00}|_{12} \langle j|_3 (\Omega_{a+b})_{23} (\chi_a)_{12} (\Omega_{a+b})^\dagger_{23} \text{tr}_4[(\chi_b)_{34}]^* |j\rangle_3 |\Phi_{00}\rangle_{12}$$
$$+ \sum_j \langle \Phi_{00}|_{12} \langle j|_3 (\Omega_{a+b})_{23} (\chi_b)_{12} (\Omega_{a+b})^\dagger_{23} \text{tr}_4[(\chi_a)_{34}]^* |j\rangle_3 |\Phi_{00}\rangle_{12}$$
$$+ \sum_j \langle \Phi_{00}|_{12} \langle j|_3 (\Omega_{a+b})_{23} (\chi_b)_{12} (\Omega_{a+b})^\dagger_{23} \text{tr}_4[(\chi_b)_{34}]^* |j\rangle_3 |\Phi_{00}\rangle_{12}$$
$$\leq \sum_{j,j'} ||\Phi_{00}\rangle||^2 ||jj'\rangle||^2 ||\Omega_{a+b}||^2 (2||\chi_a|| + ||\chi_b||)||\chi_b||$$
$$= n^2 ||\Omega_{a+b}||^2 (2||\chi_a|| + ||\chi_b||)||\chi_b||.$$

### Acknowledgements
We gratefully thank for the supports by National Key Research and development program of China (2017YFA0303700), NSFC (Grant Nos 11675113, 11434013, 11425522), "973" program (2010CB922904) and NSF of Beijing under No. KZ201810028042. Q.Q. acknowledges financial supports from NSFC (Grant No. 11704217). M.J. Zhao acknowledges financial supports from NSFC (Grant No. 11401032) and Open Foundation of State Key Laboratory of Networking and Switching Technology (Beijing University of Posts and Telecommunications) (SKLNST-20**-2-0*). G.-L.L. acknowledges support from the Center of Atomic Molecular Nanosciences, Tsinghua University, and Beijing Advanced Innovation Center for Future Chip (ICFC).

### Author Contributions
S.M.F. and W.L.Y. initiated the research project. Q.Q. established the main results. M.J.Z., H.F., S.M.F., W.L.Y. and G.L.L. joined some discussions and provided suggestions. Q.Q., M.J.Z. and S.M.F. wrote the manuscript with advice from H.F., G.L.L. and W.L.Y.

### Additional Information
**Competing Interests:** The authors declare no competing interests.

**Publisher's note:** Springer Nature remains neutral with regard to jurisdictional claims in published maps and institutional affiliations.